\documentclass[prb,twocolumn,aps,epsf,floatfix]{revtex4}
\usepackage{amssymb}
\usepackage{amsmath}
\usepackage{epsfig}

\begin{document}

\title{Andreev spectroscopy of doped HgTe quantum wells}
\author{M. Guigou and J. Cayssol}
\affiliation{CPMOH, UMR 5798, Universit\'{e} Bordeaux I,\\
33405 Talence, France}
\affiliation{}

\begin{abstract}
We investigate the Andreev reflection process in high-mobility HgTe/CdTe
quantum wells. We find that Andreev conductance probes the dynamics of
massive 2+1 Dirac fermions, and that both specular Andreev reflection and
retroreflection can be realized even in presence of a large mismatch between
the Fermi wavelengths at the two sides of the normal/superconducting
junction.
\end{abstract}

\maketitle

\section{Introduction}

Three dimensional topological insulators (TIs) and two dimensional Quantum
Spin Hall (QSH) states are novel electronic phases which are characterized
by topological invariants rather than by spontaneously broken symmetries 
\cite{ZhangToday2010,Moore2010,Hasan2010}. Both QSH states and TIs are
distinguished from ordinary insulators by the presence of conducting edge or
surface states surrounding an insulating bulk \cite%
{Kane2005,Bernevig2006,Fu2006,ZhangH2009}. These protected boundary states
were experimentally confirmed in HgTe quantum wells \cite{Konig2007,Roth2009}%
, and in three dimensional TIs like Bi$_{1-x}$Sb$_{x}$ \cite{Hsieh2008}, Bi$%
_{2}$Te$_{3}$ and Bi$_{2}$Se$_{3}$ \cite{Xia2009,Chen2009}.

All these materials follow a general mechanism whereby the strong spin-orbit
interaction drives an inversion between bands of distinct symmetry \cite%
{Bernevig2006}, e.g. opposite parities \cite{Fu2006}. Simple massive Dirac
equations describe altogether the conduction/valence and the spin degrees of
freedom of QSH systems or TIs, the crucial band inversion feature being
captured by the sign of the mass term \cite%
{ZhangToday2010,Moore2010,Hasan2010,Bernevig2006}. Therefore doped HgTe
quantum wells are also model systems to investigate the dynamics of massive
Dirac fermions in 2+1 dimensions \cite{Yokoyama2009,LBZhang2009,Novik2009}.

Combining these unique topological phases with conventional
superconductivity raises fundamental questions about topological
superconductors \cite{Qi2009,Hosur2010}, and may lead to the discovery of
novel exotic modes such as Majorana fermions \cite{Fu2008,Nilsson2008,Akhmerov2009,FuKane2009,Law2009,Linder2010} with
potential applications for spintronics and quantum computing \cite{Nayak2008}%
. Previously most studies have focused on proximity induced superconductivity within 
1D edge states  \cite{Nilsson2008,Adroguer2010} or 2D surface states  \cite{Fu2008,Akhmerov2009,FuKane2009,Law2009}. In doped TIs (resp. QSH systems), the fluctuations should be less
detrimental to superconductivity than in the 2D (or 1D) boundary states
available in the insulating regime. Indeed superconductivity was recently
reported in the topological insulator Bi$_{2}$Se$_{3}$ doped with copper
atoms \cite{Hor2010} while Heusler superconductors \cite{Chadov2010} or
Thallium based chalcogenides \cite{Yan2010} are promising candidates for a
topological superconductor.

In this context it is of primary importance to understand proximity effect
between a doped topological insulator and a s-wave singlet superconductor.
The fundamental scattering process is the Andreev reflection (AR) whereby an
incident electron is converted into a reflected hole at the interface
between a normal metal and a superconductor \cite{Andreev1964,Blonder1982}.
Moreover it was recently predicted that relativistic electrons in graphene
may experience an unusual specular Andreev reflection \cite%
{Beenakker2006,Beenakker2008,Bhattacharjee2006}.

\begin{figure}[th]
\begin{center}
\hspace*{-0.6cm} \epsfxsize7cm \epsffile{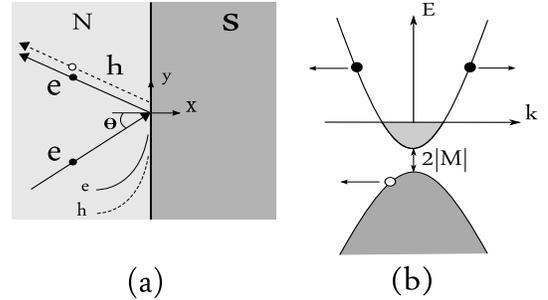}
\end{center}
\par
\vspace*{-0.7cm}
\caption{(a) Normal/superconductor (NS) contact. (b) Schematic band
structure where $2|M|$ is the bulk gap and $M$ the mass. The black dots
represent the incident and reflected electrons while the open dot stands for
the Andreev reflected hole (here interband specular reflection is shown).
Arrows represent group velocities.}
\label{fig1b}
\end{figure}

In this paper, we show that AR can be used as a probe of the carrier
dynamics in HgTe quantum wells which is intermediate between linear and
quadratic dispersion. As low gap semiconductors, HgTe wells also exhibit a
specular Andreev reflection which is sensitive to band inversion. While we
use HgTe for calculations, the general features are expected to apply for
other two-dimensional QSH systems \cite{Liu2008} and for three dimensional
TIs \cite{ZhangH2009,Xia2009,Chen2009}. Our findings also complete previous studies 
of the Andreev conductance of a NS contact in presence of Rashba spin orbital coupling \cite{Yokoyama2006}. 

The paper is organized as follows. In Sec II we introduce the model for the
NS contact. The energy and angular dependences of the Andreev reflection
processes are described in Sec. III. These single channel results are used in
Sec IV to obtain the multichannel differential conductance which is the
relevant quantity for transport experiments. 

\section{Model}

In this section, we present the model of the NS junction and describe the scattering formalism used 
to compute the Andreev probability and the differential conductance.

\subsection{Hamiltonian}

We consider a thin HgTe quantum well realized between two identical CdTe
barriers. The two-dimensional subbands of such quantum wells have been
derived from the Kane model of HgTe and CdTe using the envelope function
method \cite{Novik2005}. The QSH state is related to the crossing between
the electron-like subband $E1$ and the heavy hole-like subband $H1$ which is
controlled by the HgTe layer width \cite{Bernevig2006}. When the bulk
inversion asymmetry is neglected, those relevant bands are degenerated with
respect to the spin index ($\pm $) and have opposite parity eigenvalues.
Then the low energy dynamics of this four-band model is captured by the
massive Dirac Hamiltonian \cite{Bernevig2006} 
\begin{equation}
H(\mathbf{k})=%
\begin{pmatrix}
h(\mathbf{k}) & 0 \\ 
0 & h^{\ast }(-\mathbf{k})%
\end{pmatrix}%
,  \label{BHZ}
\end{equation}%
given in the basis order ($\left\vert E1+\right\rangle ,\left\vert
H1+\right\rangle ,\left\vert E1-\right\rangle ,\left\vert H1-\right\rangle $%
). The spin up block $h(\mathbf{k})=\varepsilon (\mathbf{k})+d_{a}(\mathbf{k}%
)\sigma _{a}$ is expressed in terms of the standard Pauli matrices $\sigma
_{a}$ ($a=1,2,3$) acting in ($\left\vert E1\right\rangle ,\left\vert
H1\right\rangle $) space. The Hamiltonian $H(\mathbf{k})$ is a Taylor
expansion with respect to the in-plane wavevector $\mathbf{k=(}k_{x},k_{y})$
whose coefficients are constrained by parity and time-reversal symmetries.
Microscopic theory further yields that $d_{a}=(Ak_{x}$,$-Ak_{y}$,$%
M(k)=M-Bk^{2})$ and $\varepsilon (\mathbf{k})=C-Dk^{2}$, where $%
k^{2}=k_{x}^{2}+k_{y}^{2}$. The parameters $A,$ $B,$ $C,$ $D$ and $M$ depend
on the quantum well geometry \cite{Bernevig2006}. In particular, the
inversion between $\left\vert E1\right\rangle $ and $\left\vert
H1\right\rangle $ is controlled by the sign of the mass term $M$, the QSH
state being realized in the inverted regime ($M<0$). The chemical potential $%
C $ determines the electronic filling of the bands which can be
electrostatically tuned by the action of a distant metallic gate.

We assume that a singlet s-wave superconductor induces a pairing potential
in some region of a ballistic HgTe/CdTe quantum well. Thereby the
Hamiltonian of Eq.(\ref{BHZ}) must be completed by particle-hole (or charge
conjugaison) symmetry leading in principle to an eight by eight
Bogoliubov-de Gennes Hamiltonian. Since superconductivity only pairs
time-reversed states, a spin up electron will be coupled with a hole in the
spin down band. Therefore electrons and holes are described by two decoupled
(four by four) Bogoliubov-de Gennes-Dirac equations like 
\begin{equation}
\begin{pmatrix}
h(-i\hbar \partial _{\mathbf{r}}) & \Delta (\mathbf{r}) \\ 
\Delta ^{\ast }(\mathbf{r}) & -h(-i\hbar \partial _{\mathbf{r}})%
\end{pmatrix}%
\Psi (\mathbf{r)}=E\Psi (\mathbf{r)},  \label{BDG}
\end{equation}%
where $\Psi (\mathbf{r)}=(\Psi _{E1+},\Psi _{H1+},\Psi _{E1-}^{\ast },\Psi
_{H1-}^{\ast })$, and the absence of magnetic field is assumed. The
excitation energy $E$ is measured from the Fermi level. The matrix $\Delta $
is diagonal at the lowest order approximation in momentum. In principle the
diagonal entries, $\Delta _{E}$ and $\Delta _{H}$, of the pairing matrix are
not equal because $\left\vert E1\right\rangle $ and $\left\vert
H1\right\rangle $ involve different combinaisons of atomic orbitals.
Nevertheless we first assume that $\Delta _{E}$ $=\Delta _{H}$=$\Delta
_{0}e^{i\phi }$ for simplicity. In a uniform system, the eigenmodes of Eq.(%
\ref{BDG}) are plane waves with momentum $\mathbf{k}$ and energy 
\begin{equation}
E(\mathbf{k})=\sqrt{(\varepsilon (\mathbf{k})\pm d(\mathbf{k}))^{2}\mathbf{+}%
\Delta _{0}^{2}},
\end{equation}%
with $d(\mathbf{k})=\sqrt{A^{2}k^{2}+(M-Bk^{2})^{2}}$. The $\pm $ sign
refers to the conduction/valence subbands.

\subsection{Scattering problem}

We now consider a straight normal/superconducting interface at $x=0$ (Fig.%
\ref{fig1b}). The normal side ($x<0$) is described by Eq.(\ref{BDG}) with $C(%
\mathbf{r})=C_{N}$ and $\Delta (\mathbf{r})=0$, while the superconducting
side may have a distinct electronic filling fixed by $C(\mathbf{r})=C_{S}$
and a uniform pairing potential $\Delta (\mathbf{r})=\Delta _{0}e^{i\phi }$.
This sharp step model is valid when the potentials vary on typical scales
smaller than the Fermi wavelength of the carriers in HgTe. We further assume
a perfect interface which is relevant for high quality contacts. Previously a NS contact in 
the presence of Rashba spin-orbit coupling was already considered in the context of spintronics \cite{Yokoyama2006}. The main distinction 
between the HgTe/CdTe effective Hamiltonian and similar Rashba systems consists in the presence the $(m-B k^2)\sigma_{3}$ term in Eqs. (\ref{BHZ},\ref{BDG}).

We solve the scattering problem with an incoming spin up quasiparticle in
the conduction band. Since the scattering is elastic and the problem
invariant by translation along $y$-axis, the excitation energy $E$ and the
transverse momentum $k_{y}$ are conserved. Hence all the scattered waves are
written hereafter as $\Psi (x)e^{ik_{y}y}$.

In the normal part, $x<0$, electrons and holes are decoupled since $\Delta (%
\mathbf{r})=0 $. The electron-like quasiparticles are described by
four-spinor plane-waves 
\begin{eqnarray}
\Psi _{e}(x)=\overset{T}{(}\chi _{\pm }(\mathbf{k}%
_{e}),0,0){e}^{ik_{ex}x},
\label{spinorelec}
\end{eqnarray}
with $\chi _{\pm }(\mathbf{k})=(\pm d(\mathbf{k}%
)+M(\mathbf{k}),A(k_{x}-ik_{y}))$ \cite{Yokoyama2009,LBZhang2009,Novik2009}.
The upperscript $T $ denotes transpose and the $\pm $ sign corresponds to
the conduction or the valence subband. For a $n$-doped HgTe well, the Fermi
level lies in the conduction subband. The dispersion equation, $%
C_{N}-Dk^{2}+ d(\mathbf{k}) =E$, allows for two possible values of $k^{2}$: 
\begin{eqnarray}
k_{1,2}^{2}(E) &=&\frac{1}{2(B^{2}-D^{2})} \big[ \gamma  \label{expressk} \\
&&\pm \sqrt{\gamma ^{2}-4(B^{2}-D^{2})(M^{2}-(C_{N}-E)^{2})} \big],  \notag
\end{eqnarray}%
where $\gamma =-A^{2}+2MB-2(C_{N}-E)D$ and $B^{2}>D^{2}$. The positive one, (%
$k_{1}^{2}>0$), corresponds to a the propagative mode in the bulk, while an
additional evanescent mode ($k_{2}^{2}<0$) shows up at interfaces \cite%
{BinZhou2008,Yokoyama2009,LBZhang2009,Novik2009}. The incident and reflected
electrons have longitidinal momentum $k_{ex}=k_{1}\cos \theta $, and $%
-k_{1}\cos \theta $ respectively, $\theta$ being the incidence angle (Fig.%
\ref{fig1b}). The evanescent electron is described by a complex momentum $%
k_{ex}=-i(-k_{2}^{2}+k_{y}^{2})^{1/2}$.

The hole-like quasiparticles are also described by the four-spinor
plane-waves 
\begin{eqnarray}
\Psi _{h}(x)=\overset{T}{(}0,0,\chi _{\pm }(\mathbf{k}_{h})){e}%
^{ik_{hx}x}.
\label{spinorhole}
\end{eqnarray}
The longitudinal wavevectors $k_{hx}$ is obtained by solving
the equation $C_{N}-Dk^{2}+ d(\mathbf{k}) =-E$ for a hole in the conduction
band (intraband AR) or $C_{N}-Dk^{2}- d(\mathbf{k}) =-E$ for a hole in the
valence band (interband AR).

In the superconducting part ($x>0$), the eigenmodes are four Bogoliubov
quasiparticles which are all evanescent below the gap ($E<\Delta _{0}$)%
\begin{eqnarray}
\Psi _{S\pm }(x) &=&\overset{T}{(}\chi _{\pm }(\mathbf{k}_{S}),\chi _{\pm }(%
\mathbf{k}_{S})e^{\pm i\beta }){e}^{(\pm ik_{Sx}-\kappa )x}, \\
\Psi _{S\pm }^{\prime }(x) &=&\overset{T}{(}\chi _{\pm }(\mathbf{k}%
_{S}^{^{\prime }}),\chi _{\pm }(\mathbf{k}_{S}^{^{\prime }}){e}^{\pm i\beta
})e^{(\pm ik_{Sx}^{\prime }\pm i\kappa )x},
\end{eqnarray}%
where $\kappa =\sqrt{\Delta _{0}^{2}-E^{2}}/\hslash v_{F}$ is an inverse
coherence length. The phase $\beta =\arccos (E/\Delta _{0})$ is
intrinsically related to electron-hole conversion at a normal
conductor-superconductor interface \cite{Andreev1964}. We have introduced
the wavevectors $k_{Sx}=(k_{1S}^{2}-k_{y}^{2})^{1/2}$ and $k_{Sx}^{\prime
}=i(k_{y}^{2}-k_{2S}^{2})^{1/2}$, where $k_{1S}^{2}$ and $k_{2S}^{2}$ are
defined by Eq.(\ref{expressk}) under the substitution $C_{N}\rightarrow
C_{S} $ and $E=0$. 
The scattering amplitudes of the four reflected modes and four evanescent
transmitted modes in the superconductor are determined by writing the
continuity of the wavefunction and of its derivative at $x=0$ (Appendix A).

\subsection{Andreev differential conductance}

When a positive bias $V$ is applied to the normal side with respect to the
superconductor, the current $I$ is carried by the injected electrons, the
reflected electrons (with amplitude $r_{ee}(E,\theta )$) and the Andreev
reflected holes (with amplitude $r_{he}(E,\theta )$). The corresponding
differential conductance of the NS interface can be written \cite%
{Blonder1982}%
\begin{eqnarray}
\frac{\partial I}{\partial V} &=&g_{0}(eV)\int dE\left( -\frac{\partial f}{%
\partial E}\right) \int\nolimits_{-\frac{\pi }{2}}^{\frac{\pi }{2}}d\theta
\cos {\theta } \nonumber\\
&&\left( 1-R(E,\theta )+R_{A}(E,\theta )\right) , 
\label{DifCond}
\end{eqnarray}%
where $g_{0}(eV)=e^{2}k_{1}(eV)W/(\pi h)$, $W$ is the width of the HgTe
quantum well along \textit{y}-axis and $f=f(E-eV)=1/(e^{(E-eV)/T}+1)$ is the
Fermi distribution of incident electrons in the normal lead N at temperature 
$T$. The probability for an electron of energy $E$ to be reflected as an electron is $R(E,\theta)=\left \vert r_{ee}(E,\theta) \right \vert ^2$ while $R_A(E,\theta)=\frac{j(E,\theta)}{j(-E,\theta)}\left \vert r_{he}(E,\theta) \right \vert ^2$ denotes the probability for Andreev reflection as a hole. The average current 
\begin{eqnarray}
j(E,\theta ) &=&-2k_{1}(E)\cos \theta \big\{(D+B)(d(\mathbf{k})+M(\mathbf{k}%
))^{2}  \notag \\
&+&A^{2}k^{2}(D-B)-A^{2}(d(\mathbf{k})+M(\mathbf{k})\big\},
\label{AverageCurrent}
\end{eqnarray}%
is derived by the standard quantum mechanical procedure (Appendix B). 

\section{Single channel Andreev reflection}

In this section, we investigate the Andreev reflection process for a single channel which is 
labelled by its energy $E$ and incidence angle $\theta$. We contrast the behaviors of the Andreev 
reflection probability $R_{A}(E,\theta )$ at high and low doping levels in the normal part of the NS junction. In the later case, the subgap Andreev reflection can be either an intraband or interband process whereas it is always intraband in heavy doped wells. In both cases, varying the ratio $B/A$ allows to turn continuously the Andreev reflection probability $R_{A}(E,\theta )$ from the one obtained in graphene/superconductor junctions to the one for standard metal/superconductor junctions.
  
\subsection{Heavy doping $%
\left\vert C_{N}\right\vert \gg \Delta _{0}$}

\begin{figure}
\hspace*{0.3cm}\includegraphics[width=7.cm]{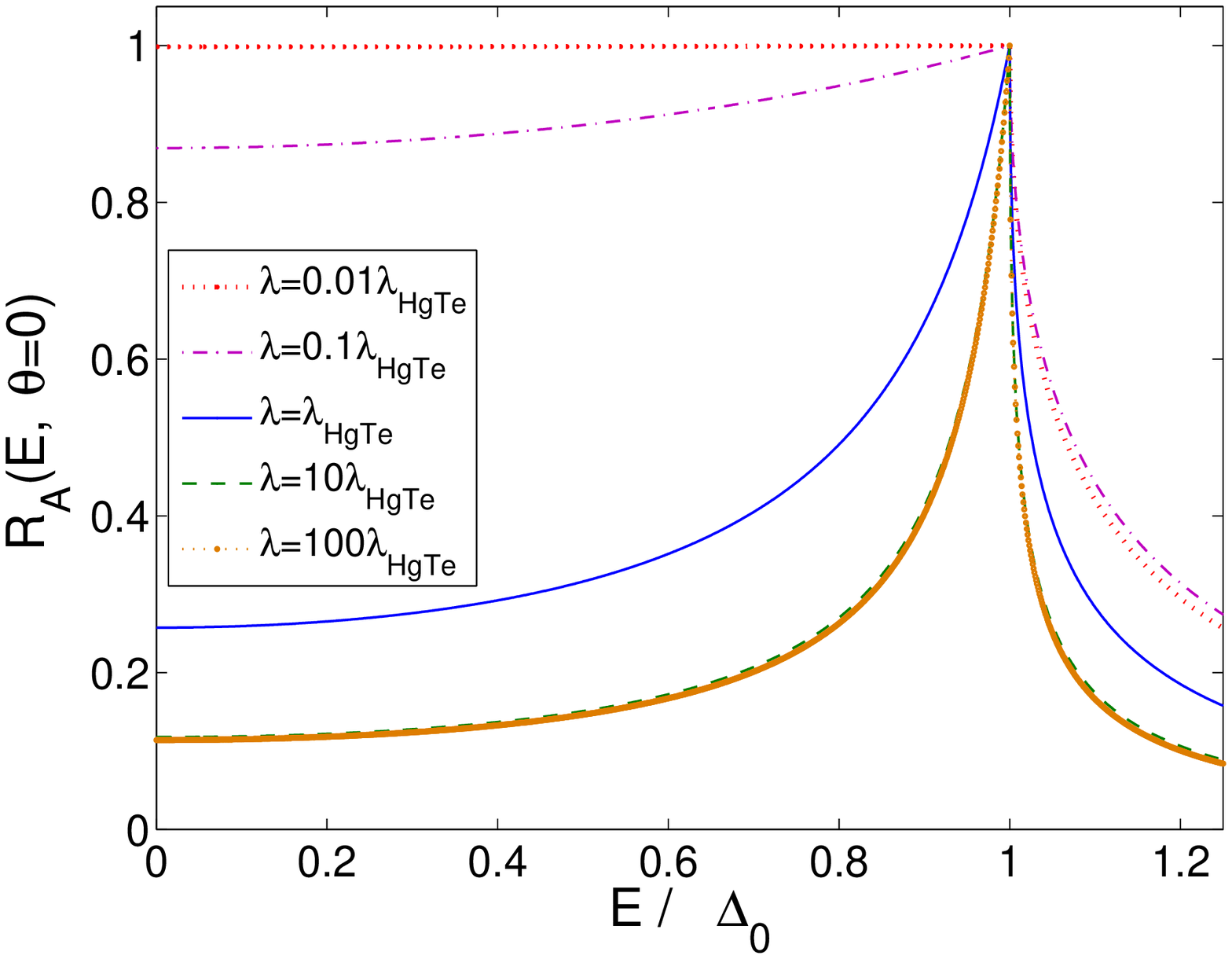}
\hspace*{0.3cm}\includegraphics[width=7.cm]{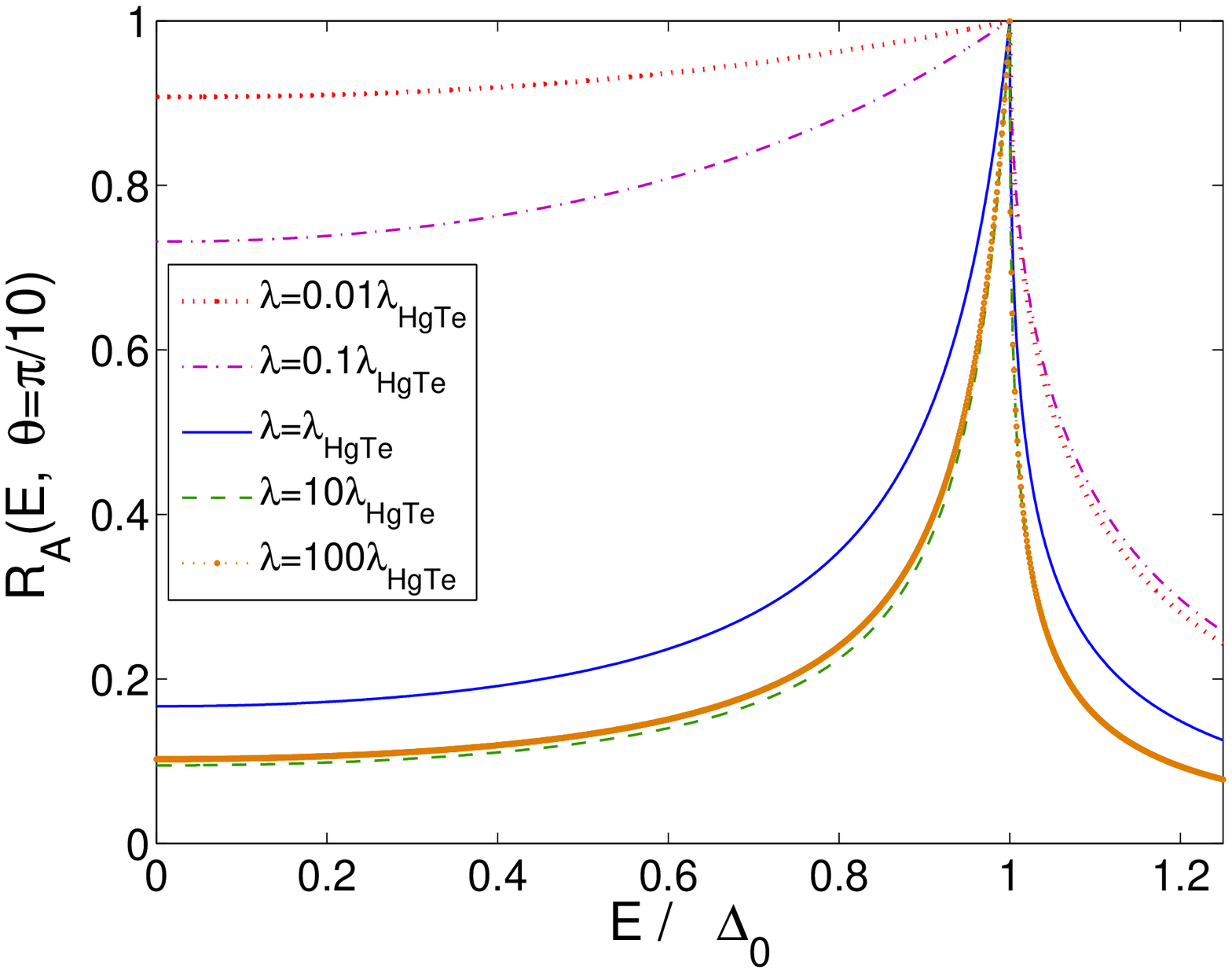}
\caption{(Color Online). Single channel Andreev reflection probability through the NS junction for $\theta=0$ (top panel) and $\theta=\pi/10$ (bottom panel). The electronic fillings are
set by $-C_{N}=30$ meV and $-C_{S}=1$ eV, while the induced gap is $\Delta
_{0}=1$ meV. All curves are plotted for $A=4$ $\mathrm{eV}\cdot \mathrm{%
\mathring{A},}$ and in the inverted regime $M=-0.1\Delta _{0}$. The Andreev
reflection probability is progressively suppressed when the quadratic dispersion is
increased: $-B=0.7$; $7$; $70$; $700$ and $7000$ $\mathrm{eV}\cdot \mathrm{%
\mathring{A}}^{2}$ from top to bottom. In contrast to $B$, the parameter $%
D$ has very little impact on Andreev reflection probability provided $%
\left\vert B \right\vert > \left\vert D \right\vert$. Hence we have set $D=0$
in order to allow for a broader window of variation for the parameter $B$.}
\label{RAheavydoping}
\end{figure}

We first consider HgTe wells with typical metallic doping, \emph{i.e.} $%
\left\vert C_{N}\right\vert \gg \Delta _{0}$. Then an incident electron (energy $E$) from the 
conduction band always finds an electron of the same band (energy $-E$) to form a Cooper pair thereby 
realizing an intraband Andreev conversion. Therefore 
the Andreev reflection is a standard retroreflection because the electron energy $E$ is always much smaller than the conduction band Fermi energy $\left\vert C_{N}\right\vert$.

We find that the probability $R_{A}(E,\theta =0)$ is strictly
monotonic (increasing) below the gap, and presents a singularity at $%
eV=\Delta _{0}$. The AR probability is controlled
by the Fermi wavelength mismatch (FWM) between the normal and
superconducting sides, and usually decreases when the FWM increases. At given electronic fillings (or FWM), the Andreev
probability is suppressed when the ratio $B/A$ is increased,
namely when going from Dirac (purely linear dispersion) to Schr\"{o}dinger
(purely quadratic dispersion) dynamics (Fig. \ref{RAheavydoping}). This crossover is
characterized by the dimensionless parameter $\lambda ={k_{1}(0)B/}{A=(n/n}%
_{0}{)}^{1/2}${,} where ${n=k_{1}^{2}(0)/(2\pi )}$ is the two-dimensional
carrier density in the conduction subband while ${n}_{0}=A^{2}/({2\pi B}%
^{2})\simeq 5.10^{12}$ cm$^{-2}$ for $A=4$ $\mathrm{eV}\cdot \mathrm{\mathring{A%
}}$ and $B=-70$ $\mathrm{eV}\cdot \mathrm{\mathring{A}}^{2}$. When electrons are injected through the junction with a finite incidence angle, the Andreev reflection probability decreases but has qualitatively the same energy dependance than the $\theta=0$ mode (Fig. \ref{RAheavydoping}). The general crossover between purely linear or quadratic dispersion still pertains for any incidence angle.

\subsection{Low doping $\left\vert
C_{N}\right\vert <\Delta _{0}$}

We now turn to the case of low doping in the normal side, namely $\left\vert
C_{N}\right\vert <\Delta _{0}$. At low voltage bias ($0<eV<\left\vert C_{N}\right\vert -\left\vert
M\right\vert )$, the reflected hole belongs to conduction band thereby
realizing the standard intraband AR which is a retroreflection. At
intermediate voltages ($\left\vert C_{N}\right\vert -\left\vert M\right\vert
<eV<\left\vert C_{N}\right\vert +\left\vert M\right\vert )$, the AR is
completely suppressed in the bulk because the hole should be in the
semiconducting gap. At higher voltages ($eV>\left\vert C_{N}\right\vert
+\left\vert M\right\vert $), the reflected hole corresponds to the removal
of a valence band electron. Then the AR is an interband process and a specular reflection. In brief, Andreev reflection allows for 
a spectroscopy of the HgTe/CdTe well 2D spectrum since for instance the energy window where the AR is totally suppressed is $2 \left\vert M\right\vert$, i.e. 
twice the semiconducting gap, for the mode $\theta=0$.

\begin{figure}
\includegraphics[width=7.5cm]{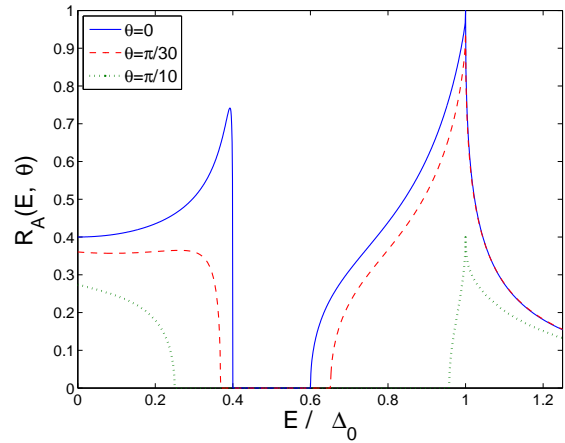}
\caption{(Color Online). Single channel Andreev reflection probability in the inverted regime ($M=-0.1\Delta_0$) at low doping ($-C_N=0.5\Delta_0$) and for three different incidence angles: $\theta=0$ (blue solid line), $\theta=\pi/30$ (red dashed line) and $\theta=\pi/10$ (green dotted line). The other parameters are taken from typical
HgTe wells: $A=4$ $\mathrm{eV}\cdot \mathrm{\mathring{A}}$, $B=-70$ $\mathrm{%
eV}\cdot \mathrm{\mathring{A}}^{2}$, $D=-50$ $\mathrm{eV}\cdot \mathrm{%
\mathring{A}}^{2}$. The superconductor is heavily doped (Fermi energy $%
-C_{S}=1$ $\mathrm{eV}$).}
\label{RAlowdoping}
\end{figure}

Moreover at the bottom of the conduction band, a singularity appears in the AR probability at $E = \left\vert C_{N}\right\vert -\left\vert
M\right\vert$  and normal incidence (blue solid curve in Fig. \ref{RAlowdoping}). Nevertheless, this abrupt behavior becomes smoother when electrons are injected with an non zero angle and the AR probability decreases.

Furthermore the energy window where the Andreev reflection is totally suppressed ($R_{A}(E,\theta )=0$) broadens while increasing incidence angle (Fig. \ref{RAlowdoping}). This phenomenon has the following explanation. Owing to translational invariance along the NS interface, the transverse momentum $k_y$ is conserved and one has to solve an effective one-dimensional scattering problem for each value of $k_y$. It turns out that a finite $k_y$ acts as a supplementary mass/gap in this 1D scattering problem. For instance, when $M=0$ the 2D system is gapless but the 1D problem (described by the dispersion relation $E(k_x,k_y)$ for defined $k_y$) acquires a gap proportionnal to the transverse momentum, as it is well known in graphene \cite{graphene}.

\section{Differential conductance}

In this section, we focus on the differential $\partial I/\partial V$ conductance of the NS contact which is obtained by angular integration over all the transverse channels. The $\partial I/\partial V$ characteristics are qualitatively distinct in the low and in the heavily doped regimes respectively, and they depend quantitatively  upon the ratio $B/A$ and the Fermi wavelength mismatch (FWM).  

\subsection{Heavy doping $%
\left\vert C_{N}\right\vert \gg \Delta _{0}$}

\begin{figure}[th]
\begin{center}
\epsfxsize7.5cm \epsffile{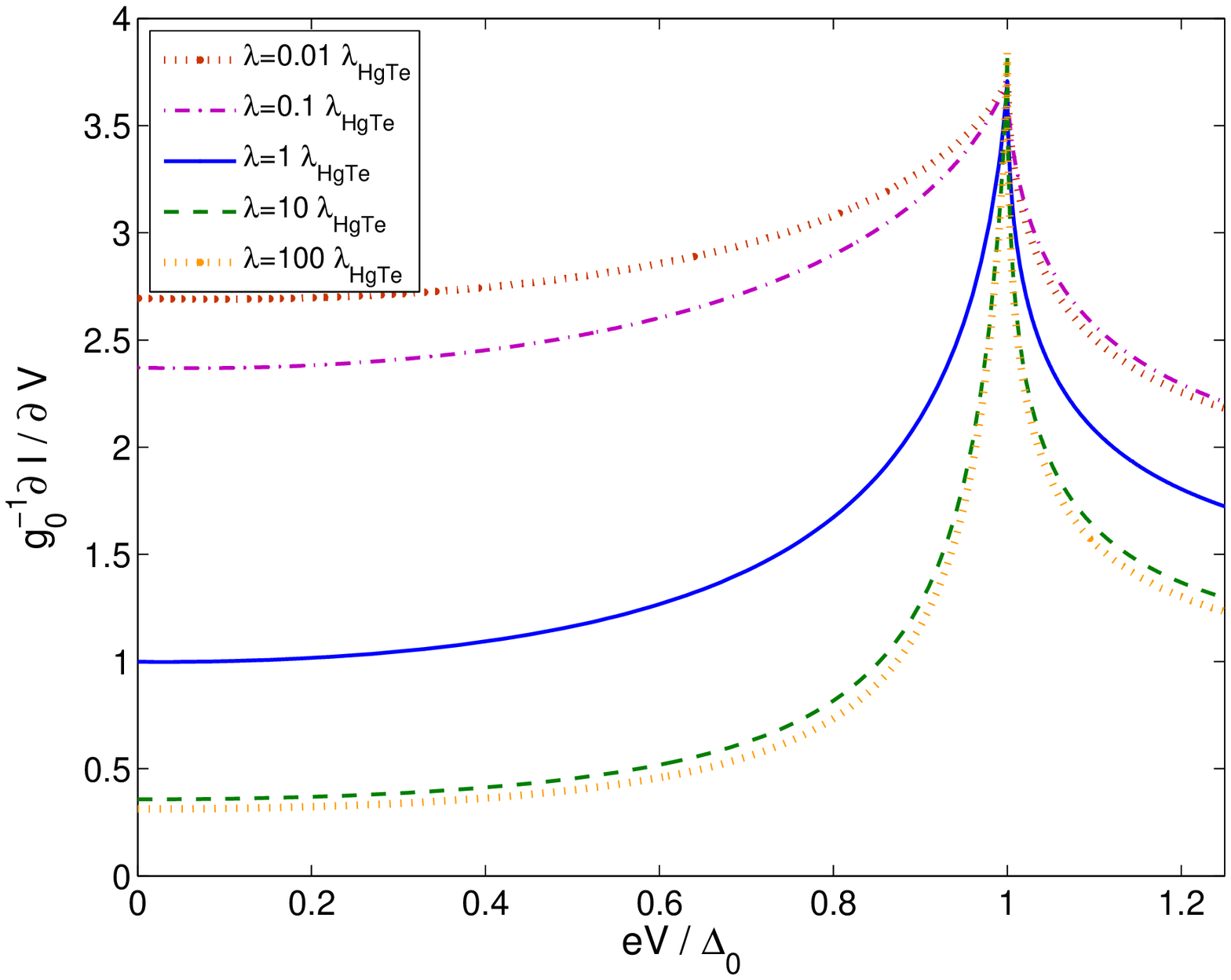}
\end{center}
\par
\vspace*{-0.7cm}
\caption{(Color online). Differential conductance of NS contact normalised
by $g_{0}(eV)=e^{2}k_{1}(eV)W/(\protect\pi h)$. Both side of the junction are heavily doped
$-C_{N}=30$ meV and $-C_{S}=1$ eV. All curves are plotted for $A=4$ $\mathrm{eV}\cdot \mathrm{%
\mathring{A},}$ and in the inverted regime $M=-0.1\Delta _{0}$. The Andreev
conductance is plotted for differents values of $B$ : $-B=0.7$; $7$; $70$; $700$ and $7000$ $\mathrm{eV}\cdot \mathrm{%
\mathring{A}}^{2}$ from top to bottom. As in Figs. \ref{RAheavydoping}, the red dotted
curve is identical to the graphene case up to a factor $2$ due to the
absence of valley degeneracy in HgTe and the blue solid curve corresponds to
typical parameters for current HgTe/CdTe. We have set $D=0$.}
\label{fig3}
\end{figure}

The voltage dependance of the differential conductance $%
\partial I/\partial V$ is inherited from the energy dependence of the AR
probability through Eq.(\ref{DifCond}). Hence, in the case of heavy doping in the N region, the $\partial I/\partial V$ characteristics are also strictly increasing 
below the gap, and present a singularity at $eV=\Delta _{0}$ (compare Fig. \ref{fig3} and Fig. \ref{RAheavydoping}). The differential conductance of
a normal potential step is asymptotically recovered far above the gap ($%
eV\gg \Delta _{0}$).  

We now discuss the crossover between Dirac (purely linear
dispersion) and Schr\"{o}dinger (purely quadratic dispersion) dynamics (Fig. \ref{fig3}). For this purpose we use again the dimensionless parameter $%
\lambda ={k_{1}(0)B/}{A}$ defined in section III.A. The increase of the ratio $B/A$ leads to the decay of the Andreev conductance. This result is in accordance with the fact that FWM strongly suppresses AR in standard metals \cite{Blonder1982}, whereas AR is very
robust against FWM at a graphene/superconductor interface \cite%
{Beenakker2006,Beenakker2008}. These contrasted behaviors can be used to
probe the relative strength of linear and quadratic dispersion in HgTe wells
(Fig. \ref{fig3}). 

Finally thin films of 3D TIs are ideal systems to probe
the complete crossover regime since their effective parameters $A$ and $B$
can be tuned by varying the film thickness \cite{Liu2008}, whereas only $M$
can be significantly varied in HgTe quantum wells.

\subsection{Low doping $\left\vert
C_{N}\right\vert <\Delta _{0}$}

In the case of low doping in the normal side ($\left\vert
C_{N}\right\vert <\Delta _{0}$), the subgap differential conductance
reveals both the semiconducting gap and the sign of the mass $M$ (Fig. \ref%
{fig2}).
Foremost the behavior of the differential conductance shows the presence of the reflected hole in the conduction/valence band through the electron-hole conversion at the interface. As a function of the voltage bias, the AR process is successively  an intraband process ($0<eV<\left\vert C_{N}\right\vert -\left\vert
M\right\vert )$, forbidden in the bulk (($\left\vert C_{N}\right\vert -\left\vert M\right\vert
<eV<\left\vert C_{N}\right\vert +\left\vert M\right\vert )$) and  an interband process ($eV>\left\vert C_{N}\right\vert
+\left\vert M\right\vert $) which is reminiscent of specular AR in graphene \cite%
{Beenakker2006,Beenakker2008}. Unfortunately the observation of specular AR
in graphene was hindered so far by disorder effects occuring at low doping.
We expect that specular AR should be more easily observed in HgTe quantum
wells since actual samples are characterized by larger elastic mean free
paths than in graphene on a substrate.

Unlike the Andreev reflection probability, the behavior of the differential conductance in the inverted regime does not emphasize any peak at energy corresponding to the bottom of the conduction band. At such energy, only one channel participates to the Andreev conductance so that its contribution is tiny compared to the number of channels that are considered in the the ballistic conductance $g_0(E)$.

\begin{figure}[th]
\begin{center}
\epsfxsize7.5cm \epsffile{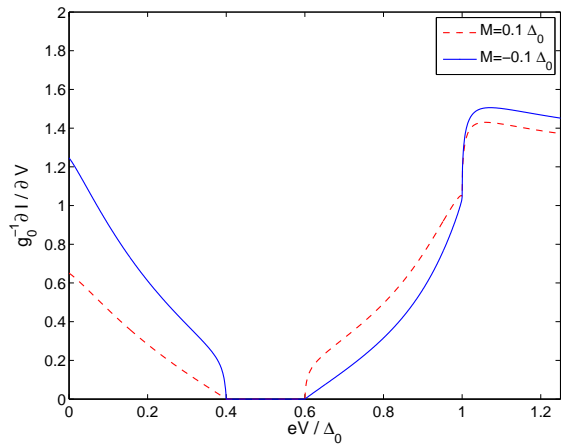}
\end{center}
\par
\vspace*{-0.7cm}
\caption{(Color online). Differential conductance in the non inverted ($M=0.1\Delta _{0}$,
dashed red line) and inverted ($M=-0.1\Delta _{0}$, solid blue line) regime
with $-C_{N}=0.5\Delta _{0}$. The parameters $A$, $B$ and $D$ are the same than in Fig. \protect\ref{RAlowdoping}.}
\label{fig2}
\end{figure}

Moreover there is a sizeable difference between the normal ($M/B<0$) and the
inverted regimes ($M/B>0$) (Fig. \ref{fig2}). Below the gap, the interband
specular AR is enhanced in the non inverted regime with respect to the
inverted regime, whereas Andreev conduction is stronger in the inverted
regime when $0<eV<\left\vert C_{N}\right\vert -\left\vert M\right\vert $
(intraband AR) or above the gap. It should be emphasized that, even at the
band crossing $M=0$, the AR in HgTe/CdTe quantum wells differ from graphene
owing to the presence of quadratic terms $Bk^{2}$ and $Dk^{2}$ in the
Hamiltonian Eq.(\ref{BDG}).

\subsection{Experimental realization}

\begin{figure}[th]
\begin{center}
\epsfxsize7.5cm \epsffile{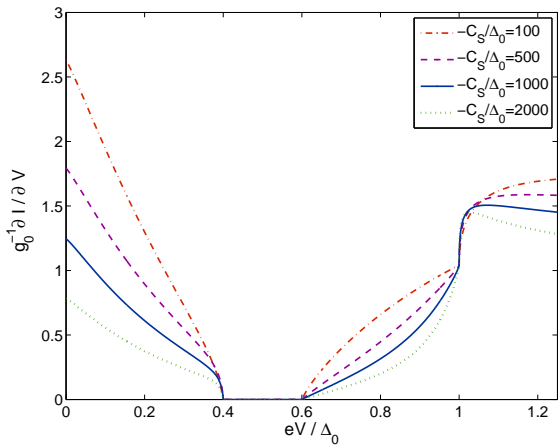}
\end{center}
\par
\vspace*{-0.7cm}
\caption{(Color online). Influence of the Fermi energy $-C_{S}$ in the superconductor while $%
-C_{N}=0.5\Delta _{0}$, $M=-0.1\Delta _{0}$. The values of the parameters $A$%
, $B$ and $D$ are the same than in Fig. \protect\ref{RAlowdoping}.}
\label{fig4}
\end{figure}

The predicted differential conductance can be checked experimentally by
realizing transparent enough contacts on existing mercury telluride
heterostructures. In particular at low doping, the specular (interband) Andreev reflection 
could be observed in HgTe/CdTe quantum wells with a superconducting contact. Nevertheless mass fluctuations, related to spatial
variations of the HgTe layer width, may hinder the observation of specular
AR in such low doped HgTe/CdTe wells (Fig. \ref{fig2}). 

Another difficulty is related to
the fact that the influence of doping in the superconducting region might be
difficult to separate from the one of other parameters, like the mass $M$ or
the ratio $B/A$, using Andreev measurements only (Fig. \ref{fig4}). Moreover $\left\vert C_{S}\right\vert $ is neither controlled nor even known in
contrast to $\left\vert C_{N}\right\vert $ which can be tuned by gating. We
anticipate that future experiments will solve this issue by combining
transport measurements through NN and NS junctions.

Finally, one should also expect in the inverted regime a contribution from the helical edge states. Since there is perfect Andreev reflection at a QSH insulator/superconductor junction, this contribution should be $4e^2/h$ ($2e^2/h$ per edge)\cite{Adroguer2010}. In the metallic regime, the large number of channels makes this edge conductance 
negligible with respect to the bulk conductance. Nevertheless near the band edges, i.e. at very low doping, the edge contribution becomes comparable with the bulk one.

\section{Conclusion}

In conclusion, Andreev conduction is a probe of the carrier dynamics in
doped HgTe/CdTe quantum wells. Indeed the underlaying AR mechanism is
extremelly sensitive to the balance between Schr\"{o}dinger (purely
quadratic dispersion) and Dirac (purely linear dispersion) dynamics. The
unavoidable FWM between the normal and superconducting sides is not so
harmful in HgTe/CdTe wells than in usual metals (but also more detrimental than in graphene). Furthermore we expect that our analysis
of intraband AR pertains for strong TIs in the doped regime, like Bi$_2$Se$%
_3 $ or Bi$_2$Te$_3$, since they are described by similar effective
Hamiltonians than HgTe/CdTe quantum wells \cite{ZhangToday2010,ZhangH2009}.
Nevertheless those large gap 3D TIs are less favorable than the small gap
HgTe/CdTe wells for the observation of specular AR.

\medskip

We thank Bjoern Trauzettel and Patrik Recher for very useful discussions. We acknowledge funding from the Agence Nationale de la Recherche under Grants
No. ANR-07-NANO-011-05 (ELEC-EPR) and ANR-10-BLAN-xxx-xx (IsoTop). J.C. also acknowledges supports from the Institut de Physique Fondamentale de Bordeaux. 

\section{Appendix}

\subsection{Boundary conditions}
We provide a derivation of the boundary conditions we used in this paper to determine the scattering amplitudes.

The Bogoliubov-de Gennes matrix equation Eq.(\ref{BDG}) consists in four scalar equations. For definiteness let us consider the first scalar equation:  
\begin{eqnarray}
&&(C+M)\Psi_{E1+}(x,y)+(D+B)(\partial_{x}^2+\partial^2_y)\Psi_{E1+}(x,y)\nonumber\\
&&-iA\partial_x\Psi_{H1+}(x,y)+A\partial_y\Psi_{H1+}(x,y)\nonumber\\
&&+\Theta(x)\Delta_0e^{i\phi}\Psi^{\ast }_{E1-}(x,y)=E\Psi_{E1+}(x,y),
\label{DifEq1}
\end{eqnarray}
where the Heaviside function $\Theta(x)$ indicates the presence of superconductivity only on the right side of the junction and where we made the substitution $k_{x,y}\rightarrow-i\partial_{x,y}$. Integrating Eq.(\ref{DifEq1}) over the small region $[-\epsilon ; \epsilon]$ around the interface, we obtain 
\begin{eqnarray}
&&(C+M)\int_{-\epsilon}^{\epsilon}dx\Psi_{E1+}(x,y)\\
&&+(D+B)\int_{-\epsilon}^{\epsilon}dx(\partial_{x}^2+\partial_y^2)\Psi_{E1+}(x,y)\nonumber\\
&&-iA\int_{-\epsilon}^{\epsilon}dx\partial_x\Psi_{H1+}(x,y)+A\partial_y\int_{-\epsilon}^{\epsilon}dx\Psi_{H1+}(x,y)\nonumber\\
&&+\Delta_0e^{i\phi}\int_{0}^{\epsilon}dx\Psi^{\ast }_{E1-}(x,y)=E\int_{-\epsilon}^{\epsilon}dx\Psi_{E1+}(x,y)\nonumber,
\end{eqnarray} 

Taking into account that the spinor $\Psi(x,y)=(\Psi_{E1+}(x,y),\Psi_{H1+}(x,y),\Psi^{\ast}_{E1-}(x,y),\Psi^{\ast }_{H1-}(x,y))$ is bounded in the region $[-\epsilon ; \epsilon]$, the limit $\epsilon \rightarrow 0$ yields the relation
\begin{eqnarray}
&&\lim_{\epsilon \rightarrow 0}(D+B)\big[\partial_{x}\Psi_{E1+}(\epsilon,y)-\partial_{x}\Psi_{E1+}(-\epsilon,y)\big]\nonumber\\
&&-\lim_{\epsilon \rightarrow 0}iA\big[\Psi_{H1+}(\epsilon,y)-\Psi_{H1+}(-\epsilon,y)\big]=0.
\end{eqnarray}
The second term of the left-hand side is equal to zero, according to the continuity of the wavefunctions (otherwise the first order derivative are not defined). Hence $\partial_x\Psi_{E1+}(x,y)$ is also continuous at $x=0$. 

Following the same procedure for the three others differential equations, we obtain the complete set of boundary conditions 
\begin{eqnarray}
\Psi(x,y)\arrowvert_{x\rightarrow 0^-}&=&\Psi(x,y)\arrowvert_{x\rightarrow 0^+}\nonumber\\
\partial_x\Psi(x,y)\arrowvert_{x\rightarrow 0^-}&=&\partial_x\Psi(x,y)\arrowvert_{x\rightarrow 0^+}.\nonumber\\
\end{eqnarray}

\subsection{Average currents}

We give here the derivation of the average current. The current operator in the $x$-direction is defined by\cite{LBZhang2009}
\begin{equation}
J_x=\frac{\partial H}{\partial k_{x}}\\
=\left( 
\begin{array}{cccc}
-2D_+k_x & A & 0 & 0 \\ 
A & -2D_-k_x & 0 & 0\\
0 & 0 & 2D_+k_x & -A \\
0 & 0 & -A & 2D_-k_x
\label{current}
\end{array}%
\right), 
\end{equation}
where $H$ is the Bogoliubov-de Gennes Hamiltonian appearing in Eq.(\ref{BDG}) and with $D_{\pm}=D \pm B$. The average current for spin-up electrons writes 
\begin{equation}
j_e(E)=\Psi^{\ast}_e(x) J_x \Psi_e(x),
\label{currentelec}
\end{equation}
where $\Psi_e(x)$ is defined in Eq.(\ref{spinorelec}). Substituing Eqs.(\ref{spinorelec},\ref{current}) into Eq.(\ref{currentelec}) leads to the expression Eq.(\ref{AverageCurrent}) of 
the average current.

The average current for a reflected hole can be obtained by following the same procedure under the substitutions $\Psi_e(x) \rightarrow \Psi_h(x)$ and $k_{ex} \rightarrow k_{hx}$. Then the average current of reflected holes is
\begin{eqnarray}
j_h(E,\theta ) &=&\mp 2k_{1}(-E)\cos \theta \big\{D_+(d(\mathbf{k})+M(\mathbf{k}%
))^{2}  \notag \\
&+&A^{2}k^{2}D_--A^{2}(d(\mathbf{k})+M(\mathbf{k})\big\},
\end{eqnarray}
where the $\mp$ sign refers to the conduction/valence band.

\end{document}